\documentclass[12pt]{article}
 \input epsf.sty 
\usepackage{amsfonts}
\usepackage{latexsym}

\begin{document} 
 
%macros 
\newcommand{\todo}[1]{{\em \small {#1}}\marginpar{$\Longleftarrow$}} 
\newcommand{\labell}[1]{\label{#1}} 
\newcommand{\bbibitem}[1]{\bibitem{#1}} 
\newcommand{\llabel}[1]{\label{#1}\marginpar{#1}}
\newcommand{\dslash}[0]{\slash{\hspace{-0.23cm}}\partial}

 % macros for the conical defect paper 
\newcommand{\sphere}[0]{{\rm S}^3} 
\newcommand{\su}[0]{{\rm SU(2)}} 
\newcommand{\so}[0]{{\rm SO(4)}} 
\newcommand{\bK}[0]{{\bf K}} 
\newcommand{\bL}[0]{{\bf L}} 
\newcommand{\bR}[0]{{\bf R}} 
\newcommand{\tK}[0]{\tilde{K}} 
\newcommand{\tL}[0]{\bar{L}} 
\newcommand{\tR}[0]{\tilde{R}}

\newcommand{\btzm}[0]{BTZ$_{\rm M}$} 
\newcommand{\ads}[1]{{\rm AdS}_{#1}} 
\newcommand{\ds}[1]{{\rm dS}_{#1}} 
\newcommand{\eds}[1]{{\rm EdS}_{#1}} 
\newcommand{\sph}[1]{{\rm S}^{#1}} 
\newcommand{\gn}[0]{G_N} 
\newcommand{\SL}[0]{{\rm SL}(2,R)} 
\newcommand{\cosm}[0]{R} 
\newcommand{\hdim}[0]{\bar{h}} 
\newcommand{\bw}[0]{\bar{w}} 
\newcommand{\bz}[0]{\bar{z}} 
\newcommand{\be}{\begin{equation}} 
\newcommand{\ee}{\end{equation}} 
\newcommand{\bea}{\begin{eqnarray}} 
\newcommand{\eea}{\end{eqnarray}} 
\newcommand{\pat}{\partial} 
\newcommand{\lp}{\lambda_+} 
\newcommand{\bx}{ {\bf x}} 
\newcommand{\bk}{{\bf k}} 
\newcommand{\bb}{{\bf b}} 
\newcommand{\BB}{{\bf B}} 
\newcommand{\tp}{\tilde{\phi}} 
\hyphenation{Min-kow-ski} 
 
%%%%%%%%%%%%%%%%%%%%%%%%%%%%%%%%%%% 
% Erich's macros: 
\renewcommand{\thepage}{\arabic{page}} 
\setcounter{page}{1} 

%\title 
\rightline{UPR-996-T, DCPT-02/33, hep-th/0205290} 
\vskip 1cm 
\centerline{\Large \bf The dual of nothing } 
\vskip 1cm 
 
\renewcommand{\thefootnote}{\fnsymbol{footnote}} 
\centerline{{\bf Vijay 
Balasubramanian${}^{1}$\footnote{vijay@endive.hep.upenn.edu} and 
Simon F. Ross${}^{2}$\footnote{S.F.Ross@durham.ac.uk}}} 
\vskip .5cm 
\centerline{${}^1$\it David Rittenhouse Laboratories, University of 
Pennsylvania,} 
\centerline{\it Philadelphia, PA 19103, USA} 
\vskip .5cm 
\centerline{${}^2$ \it Centre for Particle Theory, Department of 
Mathematical Sciences} 
\centerline{\it University of Durham, South Road, Durham DH1 3LE, U.K.} 
 
\setcounter{footnote}{0} 
\renewcommand{\thefootnote}{\arabic{footnote}} 
 
\begin{abstract} 
We consider ``bubbles of nothing'' constructed by analytically continuing black hole solutions in Anti-de Sitter space. These provide interesting examples of smooth time-dependent backgrounds which can be studied through the AdS/CFT correspondence.    Our examples include bubbles constructed from Schwarzschild-AdS, Kerr-AdS and Reissner-Nordstrom AdS.  The Schwarzschild bubble is dual to Yang-Mills theory on three dimensional de Sitter space times a circle.   We construct the boundary stress tensor of the bubble spacetime and relate it to the properties of field theory on de Sitter.
\end{abstract}

\section{Introduction}

There are many open questions in string theory, such as understanding
cosmological evolution or the information flow in black hole
formation, for which the key element is a better understanding of
dynamical spacetimes. There has recently been a surge of interest in
studying string theory in time-dependent backgrounds. Several authors
have discussed orbifold constructions giving solutions with tractable
string theory descriptions~\cite{orbifolds}. These spacetimes contain
singularities; this provides an opportunity to learn about novel
singularity-resolution mechanisms in string theory, but it also makes
these rather challenging examples. In another approach Sen has considered dynamical solutions of open string field theory with cosmological interpretations~\cite{sen}, but the corresponding spacetime solutions have not yet been understood.  Against this context, it is useful to consider simpler
spacetimes which exhibit interesting time-dependence. In~\cite{bubble1}, Aharony, Fabinger, Horowitz and Silverstein pointed out that the double analytic continuation of
Schwarzschild or Kerr spacetimes, dubbed ``bubbles of nothing'',
provide interesting examples of smooth time-dependent solutions. Since
these are vacuum solutions, they are consistent backgrounds for string
theory at least to leading order.

It would also be interesting to find time-dependent asymptotically AdS
solutions, as we could then use the AdS/CFT correspondence to relate
the time-dependence to the behaviour of the non-perturbative field
theory dual. By relating this dynamical background to the dual field
theory, it may be possible to sidestep, and get another perspective
on, some of the difficult issues associated with studying string
theory on these backgrounds, such as the possible appearance of non-local boundary 
interactions~\cite{bubble1}.

In this paper, we will extend the work of~\cite{bubble1} by
considering the double analytic continuation of black hole solutions
in AdS. These ``bubbles of nothing'' should then be related to some
state in the field theory dual to string theory on AdS$_p \times
S^q$. As we will see, this construction gives rise only to
asymptotically locally AdS spacetimes and it would be interesting to find an example asymptotic to global AdS.

We will focus on the AdS$_5 \times S^5$ case, as this corresponds to
the most well-understood field theory dual. For most of our results,
there will be an obvious extension to the AdS$_4 \times S^7$ and
AdS$_7 \times S^4$ cases. It might seem that the AdS$_3 \times S^3$
case was equally interesting, but the double analytic continuation of
the locally AdS$_3$ black hole solutions is just global
AdS$_3$.\footnote{This is related to the observation in~\cite{posen}
that the AdS soliton for $d=3$ is just global AdS$_3$.}

We begin by studying the analytic continuation of time and an angle ($t \to i \chi$, $\theta
\to i\tau$) of Schwarzschild-AdS$_5$ in section
\ref{schw}.\footnote{Related solutions were previously discussed
in~\cite{posen}.}  As in the flat space case, $\chi$ is periodically
identified, and the resulting geometry is only asymptotically locally
AdS (even though the proper length of the $\chi$ circle grows at large
distance). We find that the natural conformal boundary of this
spacetime is three dimensional de Sitter space times a circle (dS$_3 \times S^1$).  By the AdS/CFT correspondence, the Schwarzschild bubble should therefore be dual to ${\cal N} =4$ SU(N) Yang-Mills theory on dS$_3 \times S^1$.     The characteristic exponential expansion of the bulk spacetime is therefore  seen directly in the background for the field theory dual.   We provide evidence for the duality by computing the boundary stress tensor of the bubble spacetime and relating it to the expectation value of the stress tensor of Yang-Mills theory in dS$_3 \times S^1$.

In section \ref{kerr}, we consider the extension to rotating black
holes. Analytically continuing time, an angle and a rotation parameter ( $t \to i \chi$, $\theta \to
i\tau$, $a \to i\alpha$), we find that the presence of the negative
cosmological constant introduces a qualitatively new feature compared to flat space: the
metric has a coordinate singularity at a finite value of $\tau$. It
would be interesting to understand this breakdown of the metric in
more detail. However, the extension to include rotation does not
introduce any simplification: since the proper distance in all
directions grows like $r$, the spacetime is still locally
asymptotically AdS (unlike in flat space, where it was asymptotically
flat), and the spacetime the dual field theory lives in still has an
$S^1$ factor.

Finally, we consider the extension to charged black holes in AdS in
section \ref{rn}. We think of this charge as arising from angular
momentum on the $S^5$, so we consider the analytic continuation of time, an angle and the charge ($t \to i\chi$, $\theta \to i \tau$, $q \to i \varrho$), parallelling the
discussion of Kerr-AdS. These charged cases are interesting because
they have the same dS$_3 \times S^1$ conformal boundary, but there is
now an additional parameter in the solution.

In section \ref{concl}, we speculate about the interpretation of these
results from the dual field theory point of view, and outline a
program for future work. It is particularly appealing that the
time-dependence of the spacetime in these examples can be seen
directly in the background for the dual field theory.

{\it Note added:} While this paper was in preparation,~\cite{aargh}
appeared, which discusses some of the same solutions. 

\section{AdS-Schwarzschild bubbles}
\label{schw}

In this section, we consider the bubbles obtained by analytic
continuation of the AdS-Schwarzschild black hole. We will argue that
these are related to $SU(N)$ SYM on a background which includes a de
Sitter factor, and calculate the field theory stress tensor from the
asymptotics of spacetime by the counterterm subtraction procedure. 
The 5d AdS-Schwarzschild black hole has a metric
\begin{equation}
ds^2 = - (1 + {r^2 \over l^2} - {r_0^2 \over r^2}) \, dt^2 + (1 + {r^2
\over l^2} - {r_0^2 \over r^2})^{-1} \, dr^2 + r^2 \, (d\theta^2 +
\cos^2\theta \, d\Omega_2^2), \labell{adsschw}
\end{equation} 
where $d\Omega_2^2$ is the metric of the unit 2-sphere.  We can
analytically continue $t \rightarrow i\chi$ and $\theta \rightarrow
i\tau$ to obtain another vacuum solution to gravity with a negative
cosmological constant:\footnote{In the string theory context, the
solution of interest is the black hole $\times S^5$, with a constant
RR 5-form flux in both black hole and $S^5$ components. Since we
analytically continue two coordinates in the black hole, the RR 5-form
flux in this new spacetime will still be real.}
\begin{equation}
ds^2 = (1 + {r^2 \over l^2} - {r_0^2 \over r^2}) \, d\chi^2 + (1 +
{r^2 \over l^2} - {r_0^2 \over r^2})^{-1} \, dr^2 + r^2 \, (-d\tau^2 +
\cosh^2\tau \, d\Omega_2^2) .\labell{bubble}
\end{equation}
To get a smooth spacetime, we require $\chi$ to be identified with
period 
\begin{equation}
\Delta \chi = {2 \pi l^2 r_+ \over 2r_+^2 + l^2},
\labell{chiident}
\end{equation}
where $r_+$ is the minimum value of $r$, 
\begin{equation}
r^2 \geq r^2_+ = {l^2 \over 2} \left[ -1 + \sqrt{1 + {4r_0^2 \over
l^2} }\right] \, .  \labell{rbound}
\end{equation}
At any time $\tau$, at fixed large $r$ the space is the $\chi$ circle
times a 2-sphere.  As $r\rightarrow r_+$ the $\chi$ circle collapses,
but the 2-sphere approaches a finite size $r_+^2 \, \cosh^2\tau$.  This
2-sphere is the boundary of a ``bubble of nothing'' in AdS space which
contracts from infinite size at $\tau = -\infty$ to a minimum size at
$\tau = 0$ and then expands back out to infinite size as $\tau
\rightarrow \infty$.  The metric on the bubble boundary is that of 3d
de Sitter space.

At large $r$, this metric will approach AdS locally. This is not
obvious from the form of the asymptotic metric,
\begin{equation}
ds^2 \approx (1 + {r^2 \over l^2}) \, d\chi^2 + (1 + {r^2 \over
l^2})^{-1} \, dr^2 + r^2 \, (-d\tau^2 + \cosh^2\tau \, d\Omega_2^2).
\labell{timedep}
\end{equation}
However, we can relate this to the usual embedding coordinates
$X_1^2+X_2^2+X_3^3+X_4^2-T_1^2-T_2^2 = -l^2$ by
\begin{eqnarray} \label{embed1}
X_2 &=& r \cosh \tau \, \cos \theta \, \sin \phi, \\
X_3 &=& r \cosh \tau  \, \cos \theta \, \cos \phi, \nonumber \\  
X_4 &=& r \cosh \tau \sin \theta, \nonumber \\ 
 T_2 &=& r \sinh \tau, \nonumber \\
X_1 &=& (r^2 + l^2)^{1/2} \sinh \chi/l, \nonumber \\
T_1 &=&(r^2+l^2)^{1/2} \cosh \chi/l. \nonumber 
\end{eqnarray}
By contrast, the usual global AdS metric is
\begin{equation} \labell{ads}
ds^2 = -\cosh^2 \rho dt^2 + l^2 d\rho^2 + l^2 \sinh^2 \rho (d\psi^2 +
\cos^2\psi \,  d\Omega_2^2),
\end{equation}
where $-\pi/2 < \psi < \pi/2$. This is related to the embedding coordinates by
\begin{eqnarray} \label{embed2}
X_2 &=& l\sinh \rho \cos \psi \cos \theta \sin \phi, \\
X_3 &=& l \sinh \rho \cos \psi \cos \theta \cos \phi, \nonumber \\
X_4 &=& l \sinh \rho \cos \psi \sin \theta, \nonumber \\
X_1 &=& l \sinh \rho \sin \psi, \nonumber \\
T_2 &=& l \cosh \rho \sin t/l, \nonumber \\
T_1 &=& l \cosh \rho \cos t/l. \nonumber
\end{eqnarray}
Thus, the time dependent metric (\ref{timedep}) is related to the
standard global AdS coordinates (\ref{ads}) by:
\begin{equation}
r^2/l^2 = \sinh^2 \rho \cos^2 \psi - \cosh^2 \rho \sin^2 t/l, 
\end{equation}
\begin{equation}
\sinh \tau = { \cosh \rho \sin t/l \over [ \sinh^2 \rho \cos^2 \psi -
\cosh^2 \rho \sin^2 t/l ]^{1/2}}, \labell{taut}
\end{equation}
\begin{equation}
\sinh \chi/l = { \sinh \rho \sin \psi \over [ \sinh^2 \rho \cos^2 \psi
- \cosh^2 \rho \sin^2 t/l +1]^{1/2}}.  \labell{chit}
\end{equation}.

To understand the asymptotic metric, consider (\ref{timedep}) as a
coordinatization of AdS, which we will call the time-dependent AdS
coordinates. We see that these time-dependent coordinates do not even
cover the entirety of a single period in global AdS: the coordinate
patch has a boundary at $r=0$, corresponding to
\begin{equation} \labell{r0bdy}
\tanh \rho \cos \psi  = \pm \sin t/l
\end{equation}
In particular, on the asymptotic boundary of the spacetime in global
coordinates, $\rho \to \infty$, the boundary of the patch covered by
the time-dependent coordinates is the null lines $\psi = \pm t \pm
\pi/2$. We also see that in the time-dependent AdS coordinates, we
should use the full range $-\infty < \chi < \infty$. As in the usual
flat space case~\cite{bubble2}, the main effects of considering the
exact metric (\ref{bubble}) on the coordinates are twofold; the range
of $r$ is restricted to $r > r_+$ (which restricts us to a region of
AdS covered by the time-dependent coordinates), and the spacetime is
identified under $\chi \sim \chi + \Delta \chi$.

\begin{figure} 
%\vspace*{2in} 
  \begin{center} 
%   \leavevmode 
% \epsfxsize=6in 
 \epsfysize=1.5in 
   \mbox{\epsfbox{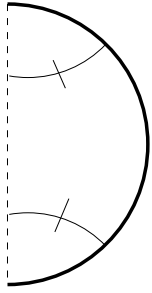}} 
\end{center} 
\caption{Periodic identification of $\chi$ in global coordinates in
the $t=0$ slice.  The figure shows the radial coordinate in AdS and
$\psi$.  Over every point in the figure there is a 2-sphere.  The
locus of points of fixed $\chi$ is shown.  }
\label{identfig} 
\end{figure}

At large distances in AdS, i.e., as $\rho \to \infty$, the restriction
to $r > r_+$ coincides with $\psi = \pm t \pm \pi/2$, the boundary of
the time-dependent coordinate patch (up to exponential corrections in
$\rho$).  The action of the periodic identification of $\chi$ on the
asymptotic metric is however slightly complicated. We will express it
in terms of the global AdS coordinates. From (\ref{chit}), we can see
that a surface $\chi = \chi_0$ in (\ref{timedep}) corresponds to
\begin{equation}
\sin \psi  = {\tanh \chi_0/l \over \tanh \rho} \cos t/l.
\labell{chi0}
\end{equation}
In, for example, the $t=0$ slice, this surface will extend to the
boundary along $\psi = \psi_0$ where $\sin \psi_0 = \tanh
\chi_0/l$. It reaches a minimum value $\rho_{min} = \chi_0/l$ at $\psi
= \pi/2$. Away from the region near the bubble, we can approximate the
bubble solution (\ref{bubble}) by the time-dependent AdS space
(\ref{timedep}), with these two restrictions. From the point of view
of the usual AdS coordinates, the periodic identification in $\chi$
will identify two surfaces of the form (\ref{chi0}), as depicted in
figure \ref{identfig}. This looks pictorially rather like the
construction of BTZ from AdS$_3$, but identifying hypersurfaces rather
than geodesics. Note however that this picture only takes into account
the effects on the coordinates, and not the fact that the bubble
geometry (\ref{bubble}) differs from the time-dependent AdS metric
(\ref{timedep}) in the interior. If we just made these identifications
on the time-dependent AdS metric (\ref{timedep}), it would not be
smooth at small $r$---in particular, in figure \ref{identfig}, it
looks like there is a finite minimum distance between the two surfaces
of fixed $\chi$, but in the true bubble geometry (\ref{bubble}), the
distance between surfaces of fixed $\chi$ goes smoothly to zero.

In the time-dependent AdS coordinates (\ref{timedep}), the natural conformal
compactification is a rescaling by $l^2/r^2$, giving a boundary metric
\begin{equation}
ds^2_\Sigma = d\chi^2 + l^2 (-d\tau^2 + \cosh^2 \tau d\Omega_2^2).
\labell{timebdy}
\end{equation}
This is a 2+1 dimensional de Sitter space times $S^1$. Thus, if we
assume the AdS/CFT correspondence can be extended to such
asymptotically locally AdS cases, we should think of the dual
description of this spacetime as given by some state of the SYM theory
on dS$_3 \times S^1$. This can be related to the usual theory on $S^3
\times R$ obtained from global AdS by considering the boundary limit
of the coordinate transformations (\ref{taut},\ref{chit}),
\begin{equation} \label{bdyxfm}
\sinh \tau = {\sin t/l \over [ \cos^2 \psi - \sin^2 t/l]^{1/2}}, \quad
\sinh \chi/l = {\sin \psi \over [\cos^2 \psi - \sin^2 t/l]^{1/2}}.
\end{equation}
These transformations take the metric (\ref{timebdy}) to 
\begin{equation} \label{bdycyl}
ds^2_\Sigma = {1 \over (\cos^2 \psi - \sin^2 t/l)} (-dt^2 + l^2
(d\psi^2 + \cos^2 \psi d\Omega_2^2)).
\end{equation}
Hence, from the boundary point of view, the coordinate transformation
between time-dependent and global AdS coordinates involves a conformal
rescaling by $\cos^2 \psi - \sin^2 t/l$. This conformal factor
vanishes at the boundary of the time-dependent AdS coordinate patch at
$\psi = \pm t \pm \pi/2$, as expected. If we also consider the effect
of the periodic identification in $\chi$, by restricting to the
fundamental region $-\Delta \chi/2 \leq \chi \leq \Delta \chi/2$, we
find that this conformal factor is non-zero except at $t = \pm \pi/2$.

From the field theory point of view, there is a single dimensionless
parameter: the ratio of the size of the $S^1$ to the radius of
curvature of the de Sitter factor. This is just $\Delta \chi/l$, and
to understand the physics from the field theory point of view, we
should express all quantities in terms of this parameter. In fact, if
we solve (\ref{chiident}) for $r_+$ in terms of $\Delta \chi$, we find
there are two roots:
\begin{equation} \labell{rplus}
r_+ = {\pi l^2 \over 2 \Delta \chi} \left[ 1 \pm \left( 1 - {2 \Delta
\chi^2 \over \pi^2 l^2} \right)^{1/2} \right].
\end{equation}
In terms of the black hole solutions, this is just the usual statement
that there is a minimum temperature for the black hole solutions, and
there are two black holes for each temperature above that value---a
smaller, unstable one and a larger stable one. 

In the discussion of the flat space analogue in~\cite{bubble1}, it was
argued that the bubble spacetime would be classically stable, but
quantum mechanically unstable. Our expectations here are slightly
different. For the larger root in (\ref{rplus}), we would expect that
the bubble will be both classically and quantum mechanically
stable. The argument for classical stability is in the same spirit
as~\cite{bubble1}: the black hole solution is classically stable, so
when one performs the analytic continuation, one expects to find no
modes of the form $e^{ik \chi}$ with negative mass$^2$ on the de
Sitter factor (it would be useful, however, to check this
explicitly). 

The quantum instability in~\cite{bubble1} was to the production of a
widely-separated bubble. First of all, we should note that the global
AdS space (\ref{ads}) with two surfaces of the form (\ref{chi0})
identified is not smooth. It is therefore not clear that we should
give the bubble of nothing the same interpretation as a
non-perturbative instability that the flat-space case had
in~\cite{bubble2}. Also, the presence of a negative cosmological
constant implies widely-separated objects cannot be treated
independently. Finally, far from the original bubble, the $\chi$
direction has a large proper radius.  As a result any identification of $\chi$ required to make a second bubble in the background of the first one will involve identifications over a very large proper length.
This also suggests that there should be no instability to creating further bubbles.

For the smaller root in (\ref{rplus}), on the other hand, there are
signs of both classical and quantum instability. The corresponding
black holes are thermodynamically unstable; it has been
argued~\cite{hubran} that this corresponds to a dynamical
instability. This may well lead to a classical instability of the bubble
solution. Also, the solution with the larger
root in (\ref{rplus}) has lower energy, so we would expect the one
with the smaller root to decay quantum-mechanically into this larger
bubble.

\subsection{The dual field theory: Stress-energy tensor}

We have shown that the asymptotic boundary of the bubble spacetime is dS$_3 \times S^1$.    Therefore, by the AdS/CFT correspondence, we expect that ${\cal N} = 4$ SU(N) Yang-Mills theory on
dS$_3 \times S^1$ should be dual to the bubble.  The time-dependence of the bubble
spacetime is reflected directly in the fact that the CFT lives on an expanding
space.     Here we will present evidence for this duality by comparing the CFT stress tensor
to the boundary stress tensor calculated from the bulk spacetime using the
counterterm subtraction procedure of~\cite{balkraus,kostas}.

Calculating the boundary stress tensor for the bubble spacetime
(\ref{bubble}) is a straightforward adaptation of the standard
calculation of the boundary stress tensor for the Schwarzschild-AdS
black hole. We must rescale the boundary stress tensor to express it
in terms of the field theory in the boundary metric
(\ref{timebdy}). The result for the bubble is
\begin{eqnarray} \labell{stress}
T^\chi_\chi &=& -{3 \over 16 \pi G l^3 } (r_0^2 + l^2/4) = -{3 N^2
\over 8\pi^2 l^4} \left(  { r_0^2 \over l^2} + {1\over 4} \right), \\
T^\tau_\tau &=& {1 \over 16 \pi G l^3} (r_0^2 + l^2/4)= { N^2
\over 8\pi^2 l^4} \left(  { r_0^2 \over l^2} + {1\over 4}  \right) , \nonumber \\
T^\theta_\theta = T^\phi_\phi &=& {1 \over 16 \pi G l^3}  (r_0^2 +
l^2/4)= { N^2
\over 8\pi^2 l^4} \left( { r_0^2 \over l^2} + {1\over 4}  \right) , \nonumber
\end{eqnarray}
where in the second equality we have used the standard relation $l^3/G
= 2N^2/\pi$ to rewrite the stress tensor in terms of field theory
quantities. It is interesting to compare this to the corresponding
result for the ordinary Schwarzschild-AdS case,
\begin{eqnarray}
T^t_t &=& -{3 \over 16 \pi G l^3} (r_0^2 + l^2/4)= -{3 N^2
\over 8\pi^2 l^4} \left( { r_0^2 \over l^2} + {1\over 4} \right), \\
T^\psi_\psi &=& {1 \over 16 \pi G l^3} l (r_0^2 + l^2/4)= { N^2
\over 8\pi^2 l^4} \left( { r_0^2 \over l^2} + {1\over 4}  \right), \nonumber \\
T^\theta_\theta = T^\phi_\phi &=& {1 \over 16 \pi G l^3} (r_0^2 +
l^2/4)= { N^2
\over 8\pi^2 l^4} \left( { r_0^2 \over l^2} + {1\over 4}  \right). \nonumber 
\end{eqnarray}

The positive sign of the $T^\tau_\tau$ component in (\ref{stress})
implies that this solution has a negative mass, while the negative
$T^\chi_\chi$ component is interpreted as a negative pressure (i.e., a
tension) along this direction.  The stress tensor is traceless, as in
Schwarzschild-AdS. This is as expected, since the boundary metric is the
product of a circle and a three-dimensional Einstein space, so the
trace anomaly vanishes.    Notice that the stress tensor has one piece that depends on the parameter $r_0$ and another that only depends on the cosmological constant.   Below we will argue that the latter can be understood in the dual field theory as an anomaly contribution, while the former depends on the state.

Now, the dual description is in terms of ${\cal N}=4$ SYM on the dS$_3
\times S^1$ spacetime (\ref{timebdy}). This spacetime is conformally
flat. We have already seen that the coordinate transformation
(\ref{bdyxfm}) takes it to the form (\ref{bdycyl}), which is conformal
to the Einstein static universe; since flat space can be conformally
embedded in the Einstein static universe, this implies that the
boundary metric (\ref{timebdy}) is conformally flat. 
Since the spacetime is conformally flat, there is a standard result for
the stress tensor~\cite{birdav}
\begin{equation} \label{ftstress}
\langle T^\mu_\nu \rangle = -{1 \over 16 \pi^2} \left( A {}^{(1)}
H^\mu_\nu + B {}^{(3)} H^\mu_\nu \right) + \tilde{T}^\mu_\nu,
\end{equation}
where $^{(1)} H^\mu_\nu$ and $^{(3)} H^\mu_\nu$ are conserved
quantities constructed from the curvature (see~\cite{birdav} for their
definitions), and $\tilde{T}^\mu_\nu$ is a traceless state-dependent
part. For
the dS$_3 \times S^1$ space, the geometrical quantities are
\begin{equation}
{}^{(1)} H^\mu_\nu = {6 \over l^4} \mbox{ diag}(-3,1,1,1)
\end{equation}
and
\begin{equation}
{}^{(3)} H^\mu_\nu = -{1 \over l^4} \mbox{ diag}(-3,1,1,1).
\end{equation}
To fix the coefficients $A$ and $B$, we compute the trace of
(\ref{ftstress}),
\begin{equation}
\langle T^\mu_\mu \rangle = -{1 \over 16 \pi^2} \left( -6A \Box R - B
(R_{\mu\nu} R^{\mu\nu} - 1/3 R^2) \right)
\end{equation}
and compare this to the conformal anomaly for ${\cal N}=4$
SYM~\cite{kostas,agmoo}
\begin{equation} 
\langle T^\mu_\mu \rangle = {(N^2-1) \over 64 \pi^2} (2 R_{\mu\nu}
R^{\mu\nu} - 2/3 R^2),
\end{equation}
which fixes $A=0$ and $B=(N^2-1)/2$. As a result, the field theory
stress tensor is 
\begin{equation}
\langle T^\mu_\nu \rangle = {(N^2-1) \over 32 \pi^2 l^4} \mbox{
diag}(-3,1,1,1) + \tilde{T}^\mu_\nu
\end{equation}
Thus, the geometrical part of the stress tensor precisely reproduces the
second term in (\ref{stress}) that is independent of the parameter $r_0$.   

It remains to relate the state-dependent part of the field theory
stress tensor to the other term in (\ref{stress}). Since we obtained
the bubble spacetime by analytic continuation from a Euclidean
solution, there is a natural vacuum state on the bulk spacetime
defined by analytic continuation from the vacuum on the Euclidean
spacetime. Similarly, there is a natural vacuum state in the field
theory defined by analytic continuation from $S^3 \times S^1$. It is
presumably this Euclidean vacuum state we should be considering.\footnote{We are grateful to Djordje Minic for discussions on this subject.}

We will defer detailed consideration of the field theory state to
future work. Here, we will simply note that to compare to the field
theory, we should rewrite the stress tensor in terms of the
dimensionless parameter $\Delta \chi/l$. The form of the stress tensor
rewritten in terms of $\Delta \chi/l$ is lengthy, so we will not give
it explicitly; it can be easily obtained using (\ref{rbound}) and
(\ref{rplus}). Note that there are two roots in (\ref{rplus}), giving
two contributions to the partition function for a given $\Delta
\chi/l$. We expect that the larger root will dominate for the field
theory state obtained from the Euclidean vacuum.

\section{Kerr-AdS bubbles}
\label{kerr}

The Schwarzschild-AdS bubble discussed above is asymptotically locally
AdS; it would be interesting to identify asymptotically AdS
solutions. In~\cite{bubble1}, the same issue for the Schwarzschild
bubble was explored by adding rotation.  Bubbles of nothing obtained by analytic continuation of Kerr spacetimes were also considered previously in~\cite{jerome}.  We will now examine the
effects of including a rotation parameter in the AdS case. We will
find that, unlike the flat space case, this fails to remove the
identification in the asymptotic region. There is also a new subtlety
which arises from the presence of a negative cosmological constant.

To simplify comparison to the flat space treatment in~\cite{bubble1},
consider first the case $d=4$. Then the metric obtained by taking $t \to i
\chi$, $a \to i \alpha$ and $\theta \to i \tau$ in the Kerr-AdS black
hole~\cite{hhr} is
\begin{eqnarray}
ds^2 &=& {\Delta_r \over \rho^2} \left[d\chi - {\alpha \over (1+ \alpha^2
l^{-2})} \cosh^2 \tau d\phi \right]^2 + {\rho^2 \over \Delta_r} dr^2 - {\rho^2
\over \Delta_\tau} d\tau^2 \\ &&+ \cosh^2 \tau {\Delta_\tau \over \rho^2}
\left[\alpha d\chi + {(r^2 - \alpha^2) \over (1+ \alpha^2 l^{-2})}
d\phi \right]^2 \nonumber
\end{eqnarray}
where 
\begin{equation}
\rho^2 = r^2 + \alpha^2 \sinh^2 \tau,
\end{equation}
\begin{equation}
\Delta_\tau = 1 - {\alpha^2 \over l^2} \sinh^2 \tau
\end{equation}
and
\begin{equation}
\Delta_r = (r^2-\alpha^2)\left(1+ {r^2 \over l^2} \right) - 2Mr.
\end{equation} 

There is a bubble at $r=r_b$, where $r_b$ is the largest root of
$\Delta_r$. But there is also now a breakdown of the metric at $\tau =
\sinh^{-1} [l/\alpha] $, where $\Delta_\tau$ vanishes. The
curvature remains finite at this point, so it may be just a coordinate
singularity. If we write $\sinh \tau = l/\alpha - \beta^2$,
the leading-order $\beta$-dependent part of the metric is
\begin{equation}
-{2 l \alpha (l^2+r^2) \over  (l^2+ \alpha^2)} d \beta^2 + \beta^2 {2  (l^2
 + \alpha^2) \over l \alpha (l^2+r^2)}\left[ \alpha d\chi + {(r^2 -
 \alpha^2) \over (1 + \alpha^2 l^{-2})} d\phi \right]^2.
\end{equation}
For any given fixed value of $r$, this looks like the Rindler-like
metric in the future light cone of a point. It therefore seems very
likely that the singularity at $\beta=0$ corresponds to a
horizon. However, a different combination of $d\chi$ and $d\phi$ is
playing the role of the hyperbolic angle in the Rindler-like coordinates
for each $r$, so it is difficult to find a coordinate transformation
that takes us through the horizon.

Similar difficulties arise in the case $d=5$.  The
analytically-continued Kerr-AdS$_5$ solution gives the bubble metric
\begin{eqnarray}
ds^2 &=& {\Delta_r \over \rho^2} \left[d\chi - {\alpha \over (1+ \alpha^2
l^{-2})} \cosh^2 \tau d\phi + {\beta \over (1+ \beta^2
l^{-2})} \sinh^2 \tau d\psi \right]^2 + {\rho^2 \over \Delta_r} dr^2 - {\rho^2
\over \Delta_\tau} d\tau^2  \nonumber \\
&&+ \cosh^2 \tau {\Delta_\tau \over \rho^2}
\left[\alpha d\chi + {(r^2 - \alpha^2) \over (1+ \alpha^2 l^{-2})}
d\phi \right]^2
- \sinh^2 \tau {\Delta_\tau \over \rho^2} 
\left[\beta d\chi + {(r^2 - \beta^2) \over (1+ \beta^2 l^{-2})} d\psi \right]^2
\nonumber \\ &&-{(1+r^2 l^{-2}) \over r^2 \rho^2} \left[ \alpha \beta
d\chi + {\beta 
(r^2 - \alpha^2) \cosh^2 \tau \over (1 + \alpha^2 l^{-2})} d\phi -
{\alpha (r^2 - \beta^2) \sinh^2 \tau \over (1+ \beta^2 l^{-2})} d\psi
\right]^2
\end{eqnarray}
where 
\begin{equation}
\rho^2 = r^2 + \alpha^2 \sinh^2 \tau - \beta^2 \cosh^2 \tau,
\end{equation}
\begin{equation}
\Delta_\tau = 1 - {\alpha^2 \over l^2} \sinh^2 \tau + {\beta^2 \over
l^2} \cosh^2 \tau,
\end{equation}
and
\begin{equation}
\Delta_r = {1 \over r^2}(r^2-\alpha^2)(r^2 - \beta^2)\left(1+{r^2
\over l^2}\right) - r_0^2.
\end{equation} 
Here, the coordinates $\phi$ and $\psi$ are angles with period $2\pi$.
If $\alpha > \beta$, there will be a breakdown of the metric where
$\Delta_\tau$ vanishes, as before. If $\beta > \alpha$, $\Delta_\tau
>0$, but we now encounter problems where $\rho=0$. 

There is still one case left, however: $\alpha=\beta$. This leads to a
considerable simplification of the metric, which becomes
\begin{eqnarray} 
ds^2 &=& {\Delta_r \over \rho^2} \left[d\chi - {\alpha \over (1+
\alpha^2 l^{-2})} (\cosh^2 \tau d\phi - \sinh^2 \tau d\psi) \right]^2
+ {\rho^2 \over \Delta_r} dr^2 - {\rho^2 \over (1+\alpha^2 l^{-2})} d\tau^2
\nonumber \\ && - { \rho^2 \over(1+\alpha^2 l^{-2})} \cosh^2 \tau
\sinh^2 \tau (d\phi - d \psi)^2 \nonumber \\ &&+{1 \over r^2} \left[
\alpha  d\chi + { \rho^2 \over (1 +
\alpha^2 l^{-2})} ( \cosh^2 \tau d\phi - \sinh^2 \tau d\psi) \right]^2,
\labell{nicekerr}
\end{eqnarray}
where 
\begin{equation}
\rho^2 = r^2 - \alpha^2
\end{equation}
and
\begin{equation}
\Delta_r = {1 \over r^2}(r^2-\alpha^2)^2 \left(1+{r^2
\over l^2}\right) - r_0^2.
\end{equation} 
In this metric, we must restrict $r$ to $r \geq r_+$, where $r_+$ is
the largest root of
\begin{equation}
(r_+^2 -\alpha^2)^2(r_+^2 + l^2) = r_0^2 l^2 r_+^2.
\end{equation}
(Note that this equation has roots for all real non-zero $\alpha, r_0,
l$.) The periodic identifications are
\begin{equation} \labell{rident}
(\chi, \phi, \psi) \sim (\chi + \Delta \chi n_1, \phi + \Delta \chi
\Omega n_1 + 2\pi n_2, \psi + \Delta \chi \Omega n_1 + 2\pi n_3),
\end{equation}
where
\begin{equation}
\Omega = - {\alpha (1+ \alpha^2 l^{-2}) \over (r_+^2 - \alpha^2)}.
\end{equation}

The surface of the bubble is at $r=r_+$. The induced metric is
\begin{eqnarray}
ds^2 &=&  - {(r_+^2 - \alpha^2) \over (1+\alpha^2 l^{-2})} d\tau^2
 - { (r_+^2 - \alpha^2) \over(1+\alpha^2 l^{-2}) } \cosh^2 \tau
\sinh^2 \tau (d\tilde{\phi} - d \tilde{\psi})^2 \\ &&+{1 \over r_+^2}
{ (r_+^2 - \alpha^2)^2 \over (1 +
\alpha^2 l^{-2})^2} ( \cosh^2 \tau d\tilde\phi - \sinh^2 \tau d\tilde
\psi)^2, \nonumber
\end{eqnarray}  
where
\begin{equation}
\tilde \phi = \phi + {\alpha (1 + \alpha^2 l^{-2}) \over (r_+^2 -
\alpha^2)} \chi, \quad
\tilde \psi = \psi + {\alpha (1 + \alpha^2 l^{-2}) \over (r_+^2 -
\alpha^2)} \chi\, .
\end{equation}
It is also useful to consider a coordinate
\begin{equation}
\bar{\phi} = \tilde{\phi} - \tilde \psi;
\end{equation}
in terms of $(\tau,\bar{\phi},\tilde \psi)$ coordinates, the metric on
the bubble is
\begin{eqnarray} \labell{rbubble2}
ds^2 &=&  - {(r_+^2 - \alpha^2) \over (1+\alpha^2 l^{-2})} d\tau^2
 - {(r_+^2 - \alpha^2)\over (1+\alpha^2 l^{-2})  } \cosh^2 \tau
\sinh^2 \tau d\bar{\phi}^2 \\ &&+{1 \over r_+^2}
{ (r_+^2 - \alpha^2)^2 \over (1 +
\alpha^2 l^{-2})^2} ( \cosh^2 \tau d\bar\phi + d\tilde
\psi)^2, \nonumber
\end{eqnarray}  
We see that the constant $\tau$ slices of the bubble are tori. Unlike
the non-rotating case, these tori are not all of finite size. The
cycle parametrised by $\tilde{\psi}$ at fixed $\tilde{\phi}$ goes to
zero size at $\tau=0$, as we can see from the first form of the
metric. More worryingly, the $g_{\bar{\phi}\bar{\phi}}$ component in
(\ref{rbubble2}) is  
\begin{equation}
{\cosh^2 \tau (r_+^2-\alpha^2) \over (1 + \alpha^2 l^{-2})^2} \left(
{(r_+^2-\alpha^2)  \over r_+^2}
\cosh^2 \tau - (1 + \alpha^2 l^{-2}) \sinh^2 \tau \right),
\end{equation}
so the cycle parametrized by $\bar{\phi}$ at fixed $\tilde{\psi}$ will
go to zero size when $\tau$ satisfies
\begin{equation}
\tanh^2 \tau = {(r_+^2 -\alpha^2) \over r^2 (1 + \alpha^2 l^{-2})},
\end{equation}
and becomes timelike for larger values of $\tau$. We will
leave the resolution of these difficulties for future work.\footnote{See
however~\cite{aargh} for a construction in higher dimensions.} 
 
As a general comment, we note that even if we had better examples,
adding rotation wouldn't remove the asymptotic identification.  In the
flat space case, proper lengths in the $\chi$ direction are
asymptotically constant, while proper lengths in the $\phi, \psi$
directions grow linearly in $r$ at large distances. Thus, the circle
in the $\chi$ direction formed by the identification (\ref{rident})
would have divergent size at large $r$ for non-zero $\Omega$. In the
anti-de Sitter case, however, proper lengths in the $\chi$ and sphere
directions all grow linearly in $r$, but this growth is removed by the
conformal rescaling to obtain a boundary metric. Hence, replacing the
identification (\ref{chiident}) by (\ref{rident}) will not eliminate
identifications in the conformal boundary; the Kerr-AdS bubble
spacetimes are still only asymptotically locally AdS.

\section{Reissner-Nordstrom AdS bubbles}
\label{rn}

Since we are interested in considering bubbles in the context of the AdS/CFT
correspondence, and hence in spacetimes which are asymptotically
AdS$_5 \times S^5$, there is another possibility to consider: we can
add angular momentum on the $S^5$. From the five-dimensional point
of view, this corresponds to considering charged black holes: a
particularly simple example is to add three equal commuting angular
momenta, which will give electrically-charged Reissner-Nordstrom AdS
black holes~\cite{cejm}. This leads to new examples with the same
asymptotic structure as in the Schwarzschild-AdS case. 

Performing the analytic continuations $t \to i\chi$, $\theta \to
i\tau$, $q \to i \varrho$ on the solution of~\cite{cejm} gives us the
bubble solution\footnote{We must analytically continue the charge so
that the resulting ten-dimensional metric is real.}
\begin{equation}
ds^2 = \left(1 + {r^2 \over l^2} - {r_0^2 \over r^2} - {\varrho^2
\over r^4}\right) \, d\chi^2 + \left(1 + {r^2 \over l^2} - {r_0^2
\over r^2} - {\varrho^2 \over r^4} \right)^{-1} \, dr^2 + r^2 \, (-
d\tau^2 + \cosh^2\tau \, d\Omega_2^2) \labell{adsrn}
\end{equation} 
with the gauge field
\begin{equation}
A_\chi = {\sqrt{3} \varrho \over 2 r^2} - {\sqrt{3} \varrho \over 2 r_+^2} \, .
\end{equation}
As in the Schwarzschild case, we need to periodically identify $\chi$
with period 
\begin{equation} \labell{qchi}
\Delta \chi = {2 \pi l^2 r_+^5 \over 2 r_+^6 + r_+^4 l^2 + 2 \varrho^2
l^2},
\end{equation}
where $r_+$ is the largest root of 
\begin{equation}
{r_+^6 \over l^2} + r_+^4 - r_0^2 r_+^2 - \varrho^2 = 0.
\end{equation}
Note that this equation has a solution for $r_+$ for all $r_0$ and
$\varrho$; as in flat space examples, the analytic continuation
of $q$ eliminates the possibility that there is no root. 

The effects of $\varrho$ in the metric are negligible at large $r$, so
the asymptotic structure of this spacetime is the same as the
uncharged case, and we get the same dS$_3 \times S^1$ metric
(\ref{timebdy}) on the conformal boundary. Here, we can think of
$\Delta \chi/l$ and $\varrho$ (which is an R charge in the CFT) as the
appropriate parameters. 

We can determine the branch structure by considering the behaviour of
$\Delta \chi$ as a function of $r_+$. For small and large $r_+$,
$\Delta \chi \to 0$. There will be a maximum where $\Delta \chi' = 0$,
which gives
\begin{equation}
2r_+^6 - r_+^4 l^2 - 10 \varrho^2 l^2 = 0.
\end{equation}
Since this equation  has only one real root, $\Delta \chi(r_+)$ has a
single maximum. Below this maximum value, there are two solutions for
$r_+$ for given $\Delta \chi$, as in the uncharged case. (Note that
this branch structure is quite different from that obtained for real
$q$.) It would be interesting to explore the stability of these
solutions as well.

Since the boundary stress tensor is independent of sub-leading terms
in the metric, it will have the same form as in the uncharged case (\ref{stress}).
%
%\begin{eqnarray} \labell{stress2}
%T^\chi_\chi &=& -{3 \over 16 \pi G l^4}  (r_0^2 l^2 + l^4/4), \\
%T^\tau_\tau &=& {1 \over 16 \pi G l^4}  (r_0^2 l^2 + l^4/4), \nonumber \\
%T^\theta_\theta = T^\phi_\phi &=& {1 \over 16 \pi G l^4} l (r_0^2 l^2 +
%l^4/4). \nonumber
%\end{eqnarray}
%
However, because (\ref{qchi}) gives us a sixth-order polynomial to
solve for $r_+$, we cannot write the stress tensor explicitly in terms
of $\Delta \chi$ and $\varrho$.

\section{Conclusions}
\label{concl}

We have begun an investigation of time-dependent bulk spacetimes in
the context of the AdS/CFT correspondence. Inspired by the work
of~\cite{bubble1}, we have considered the smooth bubble solutions
obtained from analytic continuation of bulk black hole solutions. This
gives asymptotically locally AdS spacetimes which are dual to field
theory on simple time-dependent backgrounds. The fact that the
time-dependence of the bulk spacetime shows up as time-dependence in
the background for the dual field theory is very encouraging. It
suggests that there may be interesting connections between, for
example, the notion of particle creation on the two sides of the
duality.

We focused on asymptotically AdS$_5 \times S^5$ spacetimes, dual to
${\cal N} = 4$ SYM. The extension to other cases of interest should be
straightforward. For the simplest example, the Schwarzschild-AdS$_5$
bubble, the dual field theory lives on three-dimensional de Sitter
space cross a circle. We calculated the boundary stress tensor of the bubble spacetime and showed that that it had two pieces, one which depended on the parameters of the bubble, and the other which was universal.  We showed that this universal part is reproduced by the universal anomaly contribution to the stress tensor of Yang-Mills theory on dS$_3 \times S^1$.   It will be very interesting for the future to understand the bubble parameter dependent part of the boundary stress tensor from the dual perspective.  For given radius of the circle,
there are two bubble solutions, with different values for the minimum
radius of the bubble. We expect that the smaller bubble solution
should be unstable. A careful analysis of the classical perturbations
of both these solutions, along the lines of the analysis of the AdS
solitons in~\cite{posen}, would be very useful.

We extended the construction to include angular momentum both in the
AdS factor and on the $S^5$. For angular momentum on the AdS factor,
there are new coordinate singularities which appear after analytic
continuation for generic values of the parameters. These seem to be
associated with a horizon in the bulk spacetime, but we did not
attempt to resolve this issue in detail. Angular momentum on the $S^5$
is more tractable, and leads to a structure which is very similar to
the Schwarzschild-AdS case, but with an additional parameter. Varying
this parameter provides additional opportunities to study the
behaviour of the dual field theory. We note that unlike
in~\cite{bubble1}, adding either kind of angular momentum does not
simplify the asymptotic structure of the spacetime or alter the
late-time behaviour of the bubble.

The main direction for future work is to study the properties of the
field theory on dS$_3 \times S^1$, and attempt to relate the vacuum
states on that background to the bulk spacetimes discussed here. It
would be very interesting if particle production in the bulk and on
the boundary could be related. Another open area is to attempt to find
constructions that give tractable time-dependent asymptotically AdS
solutions, which could be related to the field theory on $R^4$. 

It would also be interesting to consider the analogues of the
construction in~\cite{orbifolds}, quotienting AdS by a timelike or a
null isometry. However, since these isometries also act on the
conformal boundary, this will lead to identifications in the dual as
well, and it may be difficult to deal with the resulting backgrounds
in the field theory.

%\newpage 
 
\vspace{0.2in} {\leftline {\bf Acknowledgments}} 
We thank Jan de Boer, Djordje Minic and Asad Naqvi for discussions.
{\small V.B.} was supported by DOE grant DE-FG02-95ER40893. {\small
S.F.R.} was supported by an Advanced Fellowship from the EPSRC.  {\small S.F.R.} thanks the University of Pennsylvania for hospitality while this work was initiated.

\end{document}